\documentclass[journal]{IEEEtran}

\usepackage[pdftex]{graphicx}
\graphicspath{{../pdf/}{../jpeg/}{../png/}}
\DeclareGraphicsExtensions{.pdf,.jpeg,.png}

\usepackage{amsmath}
\usepackage{array}
\usepackage{standalone}
\usepackage{etoolbox}
\newtoggle{arXiv}
\newtoggle{bibrun}
\usepackage{newtxtext, newtxmath}
\usepackage{graphicx}
\usepackage{color, calc}
\usepackage{array, booktabs}
\newcolumntype{L}{>{$}l<{$}}
\newcolumntype{C}{>{$}c<{$}}
\newcolumntype{R}{>{$}r<{$}}
\usepackage{tikz, tikz-3dplot}
\usetikzlibrary{arrows, arrows.meta, calc, positioning}
\usetikzlibrary{decorations.pathmorphing}	
\tikzset{>=Stealth}
\usepackage{pgfplots}
\usepackage{siunitx, physics}
\newcommand{\Exp}[1]{\mathrm{e}^{#1}}
\usepackage{enumitem}
\setlist[description]{labelindent=0pt, leftmargin=\parindent, font=\normalfont\itshape}
\usepackage{url}
\usepackage{subfigure}
\usepackage{textcomp}
\usepackage{eurosym}

\begin{document}

\title{A Fast 0.5~T Prepolarizer Module for Preclinical Magnetic Resonance Imaging}

\author{\IEEEauthorblockN{
		Juan~P.~Rigla\IEEEauthorrefmark{1},
		Jose~Borreguero\IEEEauthorrefmark{2},
		Carlos~Gramage\IEEEauthorrefmark{2},
		Eduardo~Pall\'as\IEEEauthorrefmark{2},
		Jos\'e~M.~Gonz\'alez\IEEEauthorrefmark{1},
		Rub\'en~Bosch\IEEEauthorrefmark{2},
		Jos\'e~M.~Algar\'{\i}n\IEEEauthorrefmark{2},
		Juan V. Sanchez-Andres\IEEEauthorrefmark{3},
		Fernando~Galve\IEEEauthorrefmark{2},
		Daniel~Grau-Ruiz\IEEEauthorrefmark{1},
		Rub\'en~Pellicer\IEEEauthorrefmark{2},
		Alfonso~R\'ios\IEEEauthorrefmark{1},
		Jos\'e~M.~Benlloch\IEEEauthorrefmark{2}, and
		Joseba~Alonso\IEEEauthorrefmark{2}}
	
	\IEEEauthorblockA{\IEEEauthorrefmark{1}Tesoro Imaging S.L., 46022 Valencia, Spain}\\
	\IEEEauthorblockA{\IEEEauthorrefmark{2}MRILab, Institute for Molecular Imaging and Instrumentation (i3M), Spanish National Research Council (CSIC) and Universitat Polit\`ecnica de Val\`encia (UPV), 46022 Valencia, Spain}\\
	\IEEEauthorblockA{\IEEEauthorrefmark{3}Dept. Medicine, Univ. Jaume I, 12071 Castellon, Spain}\\
	
\thanks{Corresponding author: J. Alonso (joseba.alonso@i3m.upv.es).}}

\maketitle

\IEEEtitleabstractindextext{%
\begin{abstract}
We present a magnet and high power electronics for Prepolarized Magnetic Resonance Imaging (PMRI) in a home-made, special-purpose preclinical system designed for simultaneous visualization of hard and soft biological tissues. PMRI boosts the signal-to-noise ratio (SNR) by means of a long and strong magnetic pulse which must be rapidly switched off prior to the imaging pulse sequence, in timescales shorter than the spin relaxation (or $T_1$) time of the sample. We have operated the prepolarizer at up to 0.5~T and demonstrated enhanced magnetization, image SNR and tissue contrast with PMRI of tap water, an \emph{ex vivo} mouse brain and food samples. These have $T_1$ times ranging from hundreds of milli-seconds to single seconds, while the preliminary high-power electronics setup employed in this work can switch off the prepolarization field in tens of milli-seconds. In order to make this system suitable for solid-state matter and hard tissues, which feature $T_1$ times as short as 10~ms, we are developing new electronics which can cut switching times to $\sim\SI{300}{\micro s}$. This does not require changes in the prepolarizer module, opening the door to the first experimental demonstration of PMRI on hard biological tissues.
\end{abstract}

}

\maketitle

\IEEEdisplaynontitleabstractindextext

\section{Introduction}
\IEEEPARstart{P}{repolarization} in Magnetic Resonance Imaging (MRI) refers to the enhancement of the sample magnetization by means of an intense magnetic field pulsed before the imaging sequence \cite{Macovski1993,Morgan1996,Kegler2007,Lee2005,Obungoloch2018}. This technique can boost the signal to noise ratio (SNR) in the context of low field MRI, which is pursued as an affordable alternative to standard MRI at a fraction of the cost of high-field clinical scanners \cite{Sarracanie2015,Marques2019,Sarracanie2020}.

The resonant (Larmor) frequency of a spin particle in a magnetic field of strength $B_0$ is $\omega_0=\gamma B_0$, where the proportionality constant $\gamma$ is the gyromagnetic factor. In MRI the intensity of the detected signal is proportional to $\omega_0$, as a consequence of the Faraday Law that describes the coupling between the sample spins and the inductive detectors. It is also proportional to the sample magnetization $M_0$. On the other hand, there are three main contributions to the noise variance \cite{Harpen1987}. One is due to thermal effects in the detector (resistive losses) and goes as $\omega_0^{1/2}$, another originates at the sample (dielectric losses) and is proportional to $\omega_0^2$, and the last one arises from non-ideal electronics in the reception chain and will be neglected in this analysis. All in all, the SNR is hence given by \cite{BkHaacke}
\begin{equation}\label{eq:snr}
	\text{SNR} = \kappa \frac{\omega_0 M_0}{\sqrt{c_\text{det} \omega_0^{1/2} + c_\text{sam}\omega_0^2}},
\end{equation}
where $\kappa$ is a proportionality factor which depends on multiple elements (e.g. the detector quality factor, the pulse sequence or the physical dimensions of the hardware and sample), and $c_\text{det}$ ($c_\text{sam}$) determines the weight of the detector (sample) contribution, which tends to dominate at low (high) Larmor frequencies. Without prepolarization, $M_0$ is directly proportional to the main magnetic field $B_0$, so the SNR increases monotonically with $\omega_0$, which explains the trend to explore increasingly higher field strengths in MRI \cite{Kraff2015}. Prepolarization alleviates the SNR decrease at low fields, by applying a field of strength $B_\text{p}$, so $\text{SNR}\propto |\vec{B}_\text{p}+\vec{B}_0|$.

The advantages of Prepolarized MRI (PMRI) have been already demonstrated for human \emph{in vivo} imaging \cite{Matter2006,Venook2006}, they have enabled human brain imaging at milli-tesla \cite{Savukov2013} and micro-tesla fields \cite{Inglis2013}, and PMRI has been shown to be highly efficient when combined with other magnetization enhancement techniques such as hyperpolarization \cite{Coffey2013}. However, we are not aware of PMRI applied to hard tissues or solid state matter. This is challenging due to their short spin relaxation times ($T_1$) at low fields, since the prepolarization field must be ramped down in times $\tau \lesssim T_1$ to preserve the boost in magnetization for the imaging sequence, but prepolarizers are typically high inductance to achieve large values of $B_\text{p}$.

In this paper, we present the design and performance of a 0.5~T prepolarization module which can be integrated into our ``DentMRI - Gen I'' scanner \cite{Algarin2020}. The latter is a preclinical dental system with a field of view of only 10~mm in diameter when the prepolarizer is installed, so strong prepolarization fields are possible even with a low inductance ($\approx\SI{600}{\micro H}$) coil. This allows for sub-milli-second switching of the field, which is in principle short enough for prepolarization of teeth (the hardest human tissues), with apparent $T_1$ of 13~ms for the heterogeneous sample including dentin and enamel, which we have measured at $B_0\approx\SI{0.26}{T}$. As a proof of concept,  we demonstrate switching times of $\approx\SI{30}{ms}$, fast enough for efficient prepolarization of relevant hard tissues such as bone and tendons, with a basic high power electronics module available in our laboratory. Finally, we show the first images of a mouse brain taken with this system, and demonstrate tissue contrast with PMRI with a heterogeneous sample.

\section{Prepolarizer design}
\subsection{Magnetic, mechanical and thermal design}

\begin{figure*}
	\centering
	\includegraphics[width=2.\columnwidth]{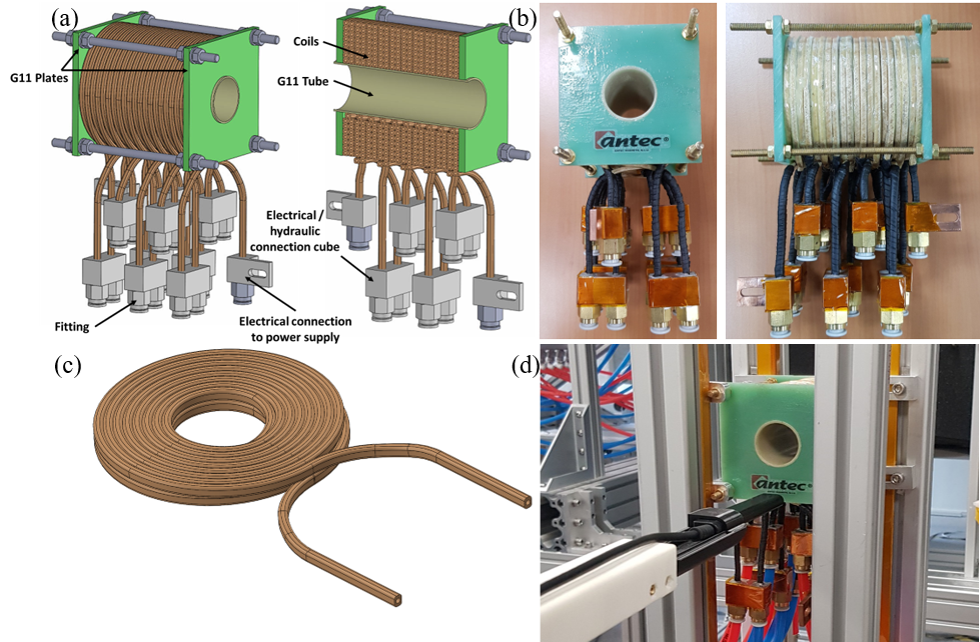}
	\caption{(a) 3D drawing of the assembled prepolarizer, including the coils and the electrical and hydraulic connections, and cross section with coil details. (b) Photographs of the prepolarizer magnet, before installation. (c) 3D drawing of a single coil layer, including two spirals. (d) Prepolarizer connectorized and assembled in the test bench for field mapping, where the Hall probe is also shown.}
	\label{fig:CadPics}
\end{figure*}

The main constraints we considered for the design of the prepolarizer were: i) it must be a module that can be easily installed into and removed from the ``DentMRI - Gen I'' system \cite{Algarin2020}; ii) it must produce a prepolarization field significantly higher than the evolution field $B_0\approx\SI{0.26}{T}$; and iii) it must be possible to ramp the field up and down in less than 10~ms to enable PMRI of hard biological tissues.

The main field in ``Gen I'' is provided by a C-shaped permanent magnet with a gap between poles of $\approx\SI{210}{mm}$. With the stack of planar gradient coils installed, the gap reduces to $\approx\SI{120}{mm}$, placing a hard boundary on the prepolarizer size and, consequently, to the maximum achievable coil inductance. After iterating through various configurations, we converged to the design shown in Figs.~\ref{fig:CadPics}(a)-(b), with a stack of ten layers such as the one in Fig.~\ref{fig:CadPics}(c), for a total length close to 90~mm. Every layer contains two spirals of inner (outer) diameter 35~mm (99~mm), each with eight windings. The conductor is manufactured by ANTEC Magnets SLU from hollow OF-OK copper tubing of section $4\times3$~mm$^2$ and an inner hole of 2~mm in diameter. This high purity copper is immune against hydrogen embrittlement, and the hollow section allows for water cooling (see below). The loops are electrically connected in series and the water cooling paths are connected in parallel to maximize heat transfer. The copper tubing is electrically isolated by a layer of fiberglass with polyester (Fig.~\ref{fig:CadPics}(b)). In this way, we prevent short-circuits between the windings and layers, and we increase mechanical stability against stress due to the Lorentz forces between the prepolarization and main magnets. The loops are fixed using a structure based on G11 fiberglass epoxy laminate. The electrical connections between the different loops are done through copper cubes, which accommodate fittings for the cooling tubing. The total cost for the module is approximately 7,500~\euro.

\begin{figure}
	\centering
	\includegraphics[width=0.8\columnwidth]{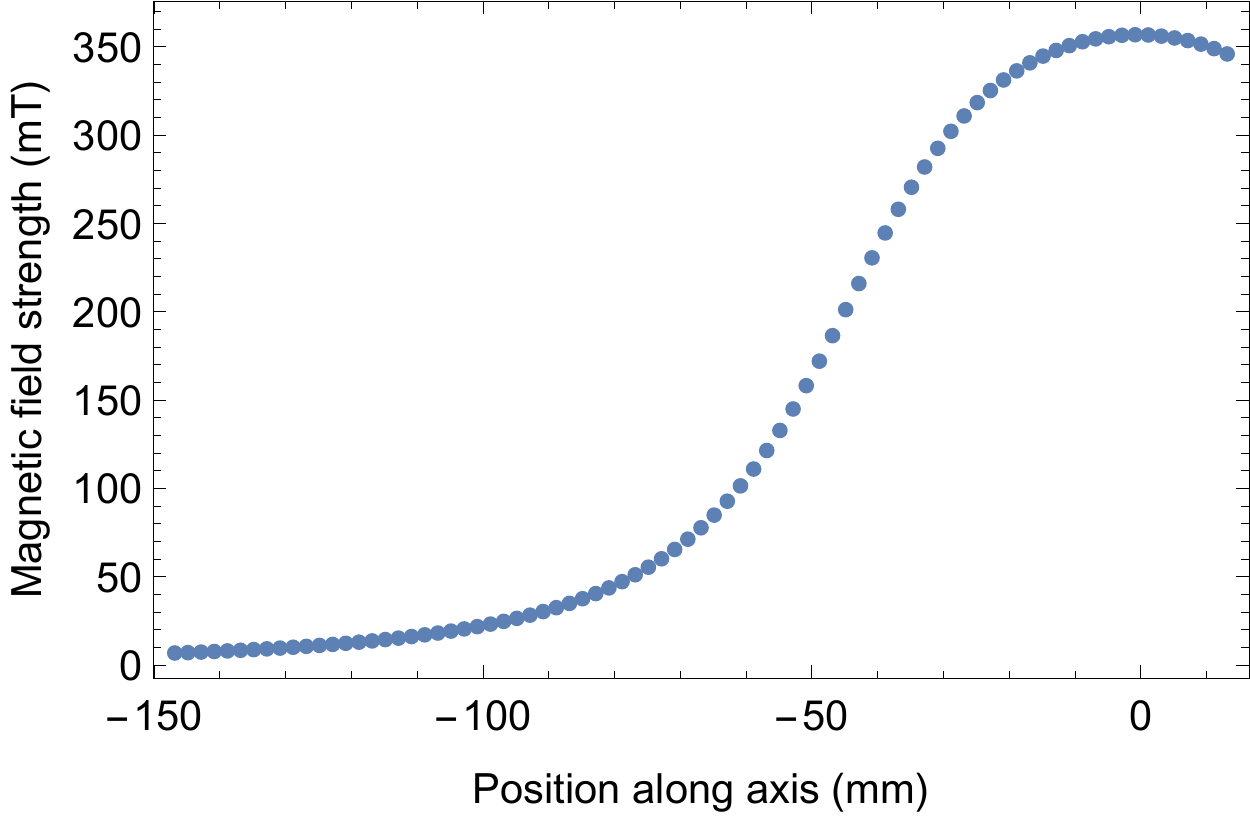}
	\caption{Magnetic-field profile measured along the $x$-axis for $I_\text{p} \approx 190$~A. The zero in the abscissa corresponds roughly to the solenoid center.}
	\label{fig:BProfile}
\end{figure}

\begin{table}
\caption{Prepolarizer coil parameters}
\centering
\begin{tabular}{c c}
\hline
Parameter & Value\\
\hline
Inductance ($L$) & \SI{600}{\micro H}\\
Resistance ($R$) & \SI{75}{m \ohm} \\
Coil efficiency ($\eta$) & \SI{1.9}{mT/A}\\
\hline
\end{tabular}
\label{tab:Prepol}
\end{table}

All in all, the coil has a self inductance $L\approx\SI{600}{\micro H}$ and a dc resistance $R\approx\SI{75}{m \ohm}$ (Tab.~\ref{tab:Prepol}). The prepolarizing field reaches $B_\text{p}=\SI{0.48}{T}$ (at the center of the solenoid) for a drive current $I_\text{p}\approx\SI{255}{A}$ (coil efficiency $\eta\approx\SI{1.9}{mT/A}$), decreasing by around 30~\% when measured 4~cm away from the center along the axis. This level of homogeneity is perfectly tolerable in PMRI, since its only effect is to shade the resulting images where the field is weaker \cite{Macovski1993}. The field strength profile measured along the coil axis is shown in Fig.~\ref{fig:BProfile} for $I_\text{p}\approx\SI{190}{A}$, the maximum intensity available from the Danfysik 9100 power supply employed for these tests. For this, we use a Hall-probe magnetometer (Metrolab THM1176-MF) mounted on a home-made 3D positioning system (Fig.~\ref{fig:CadPics}(d)).

\begin{figure}
	\centering
	\includegraphics[width=1.\columnwidth]{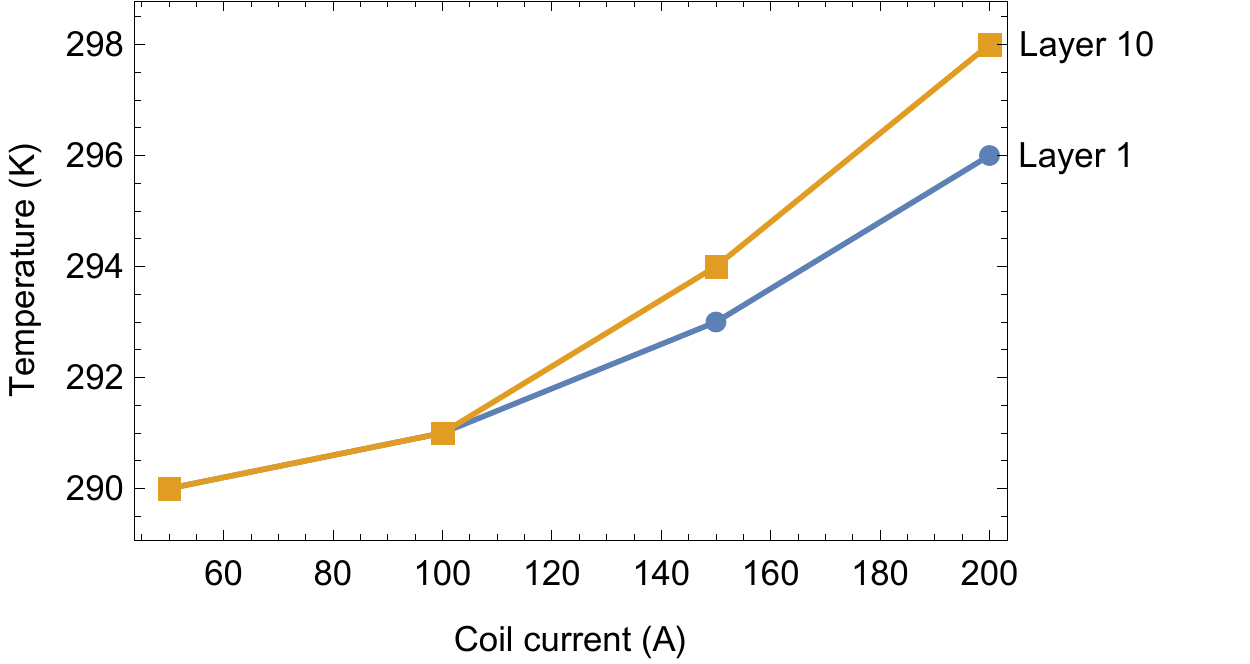}
	\caption{Temperature of the first and last layers of the prepolarizer as a function of the coil current $I_\text{p}$, measured after reaching thermal equilibrium.}
	\label{fig:Thermal}
\end{figure}

The intense currents required to achieve the desired values of $B_\text{p}$ impose the use of cooling mechanisms to remove the heat dissipated by Joule effect. At $I_\text{p}=\SI{255}{A}$ ($B_\text{p}\approx\SI{0.48}{T}$), the dc resistance of $\approx\SI{75}{m\ohm}$ means the dissipated power is expected to be $\approx\SI{4.9}{kW}$. Thermal simulations in Comsol indicate that a water flow of $\approx\SI{13}{l/min}$ and a pressure drop of $\approx\SI{4}{bar}$ can limit the temperature increase at the coils to <13~K at a duty cycle of 100~\%. To this end, we employ a water chiller (SMC HRS090-AF-40), which feeds in parallel all ten layers in the stack that forms the prepolarizer. The experimental characterization of the thermal behavior was performed at a maximum $I_\text{p}$ of 190~A, for which a jump of 6~K (8~K) is measured at the first (last) layer (Fig.~\ref{fig:Thermal}). The chiller can take up to 10~kW of power, so this system should be able to run stably at >360~A ($B_\text{p}>\SI{0.68}{T}$), although we have never tested it beyond 270~A.

\subsection{Electronics design}\label{sec:elec}
In the context of the Histo-MRI consortium, we have built a scanner for high resolution imaging where the gradient coils ($\approx\SI{10}{\micro H}$) can be switched on and off in less than $\SI{10}{\micro s}$ \cite{Rigla2020,Grau-Ruiz2020,HISTOMRI}. As a result of this project, Danfysik A/S (Taastrup, Denmark) will have developed fast Pulsed Power Supply Units (PPSU), which can deliver up to 500~V and 500~A. Once these units are tested and shipped to our laboratories, minimal software updates can make them suitable for the $\approx\SI{600}{\micro H}$ load from the prepolarizer. This should allow for rise/fall times $\tau=L I_\text{p} / V\approx\SI{300}{\micro s}$ (see Tab.~\ref{tab:Pulses}), well below the $T_1$ of the hardest biological tissues.

\begin{figure}
	\centering
	\includegraphics[width=1.\columnwidth]{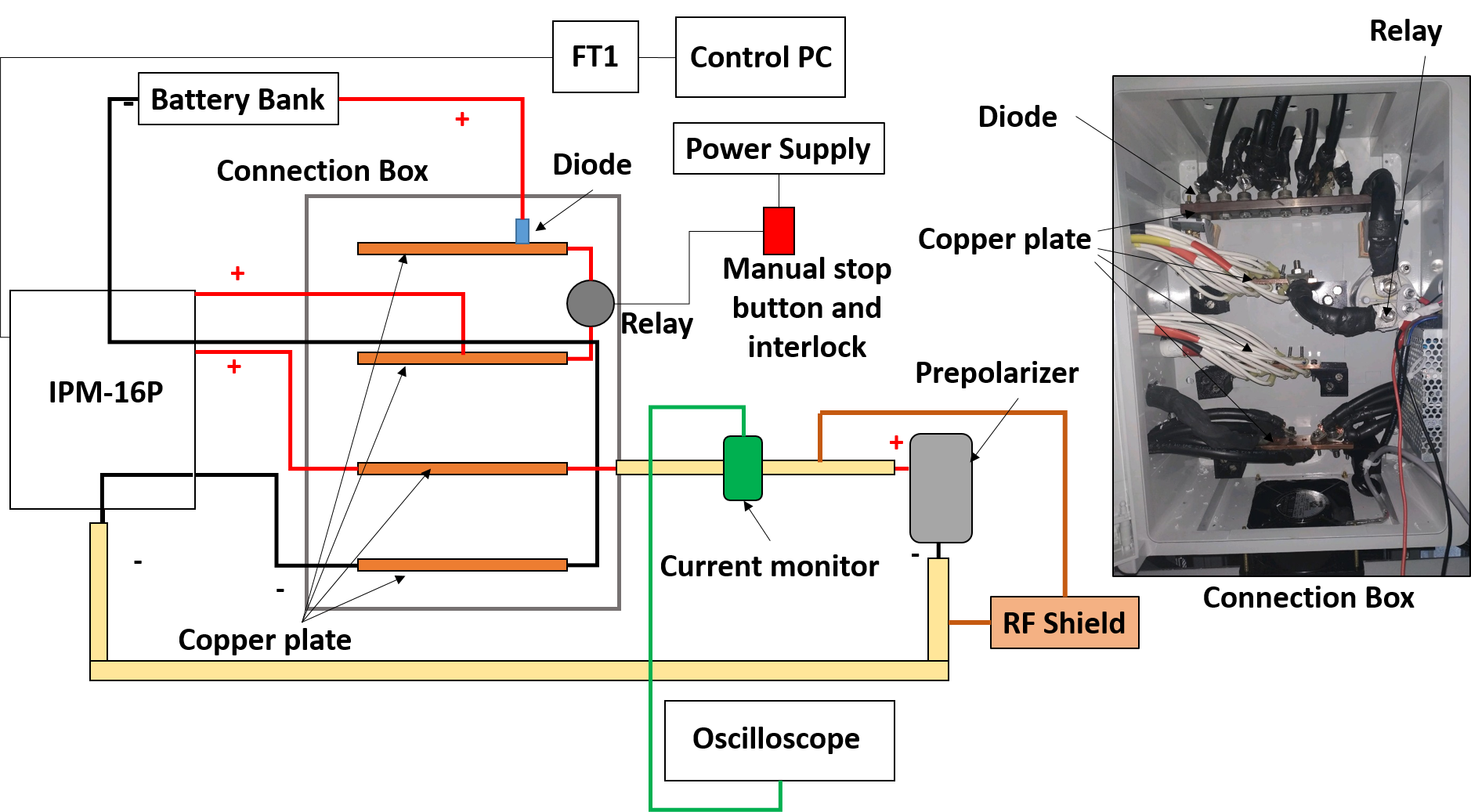}
	\caption{High power electronics equipment used for demonstrating PMRI, including the IPM-16P for high power switching, the battery bank and connection box for supplying the IPM-16P, the control PC and FT1 to trigger the IPM-16P, and the prepolarizer magnet and measurement devices. A photograph of the inside of the connection box is also shown for reference.}
	\label{fig:BatterySetup}
\end{figure}

In the meantime, we have used high power electronics equipment present in the laboratory (see Fig.~\ref{fig:BatterySetup}) for demonstrating PMRI of softer tissues, which is less demanding in terms of the switch off time $\tau$. This setup relies on a high power switching module (IPM-16P from Eagle Harbor Technologies Inc.) based on 16 Insulated-Gate Bipolar Transistors (IGBTs). Each of these can take 1,200~V and 80~A, for a total device current of 1,280~A. However, the IPM-16P is driven from a battery bank consisting of 16 commercial batteries (12~V, LFS 105N from Varta Automotive). For our total load resistance of $\approx\SI{95}{m\ohm}$ ($\approx\SI{75}{m\ohm}$ from the coil, $\approx\SI{20}{m\ohm}$ from cables and connections), every battery outputs a current $\approx\SI{128}{A}$ when fully charged. We therefore arrange the batteries in pairs, each pair with both batteries connected in series. In order to prolong the batteries' discharge time, we use all 8 pairs simultaneously, connected in parallel. The positive end of every pair is connected through a diode (VS-70HF40) to a thick copper plate that serves as the positive input of a custom-built high-power connection box (see Fig.~\ref{fig:BatterySetup}, right). This is then connected to a second copper plate (positive output from the connections box) through a relay (Kilovac LEV200), enabled by a manual stop button and an interlock to ensure the prepolarizer is never charged if the water chiller is not functional. The negative ends of the battery pairs are similarly connected (without diodes) to the bottom copper plate, which serves as both negative input and output. The IPM-16P outputs are connected to the prepolarizer with cables of section \SI{95}{mm^2}, which we shield with a conducting mesh (Scotch electrical shielding tape 24). This mesh is grounded through the radio-frequency (rf) shield of the resonant coil used for coherent spin manipulation and resonant MRI signal detection. This strongly suppresses rf noise pickup and 50~Hz inductive couplings which otherwise strongly deteriorate our MRI signals (see Sec.~\ref{sec:noise}). The control system is based on a RadioProcessor-G console from SpinCore Technologies Inc. and is described elsewhere \cite{Algarin2020}. To control the prepolarizer via the console, we use a TTL line which triggers the IPM-16P switches after optical decoupling in the FT1 module (Eagle Harbor Technologies Inc.).

\begin{figure}
	\centering
	\includegraphics[width=1.\columnwidth]{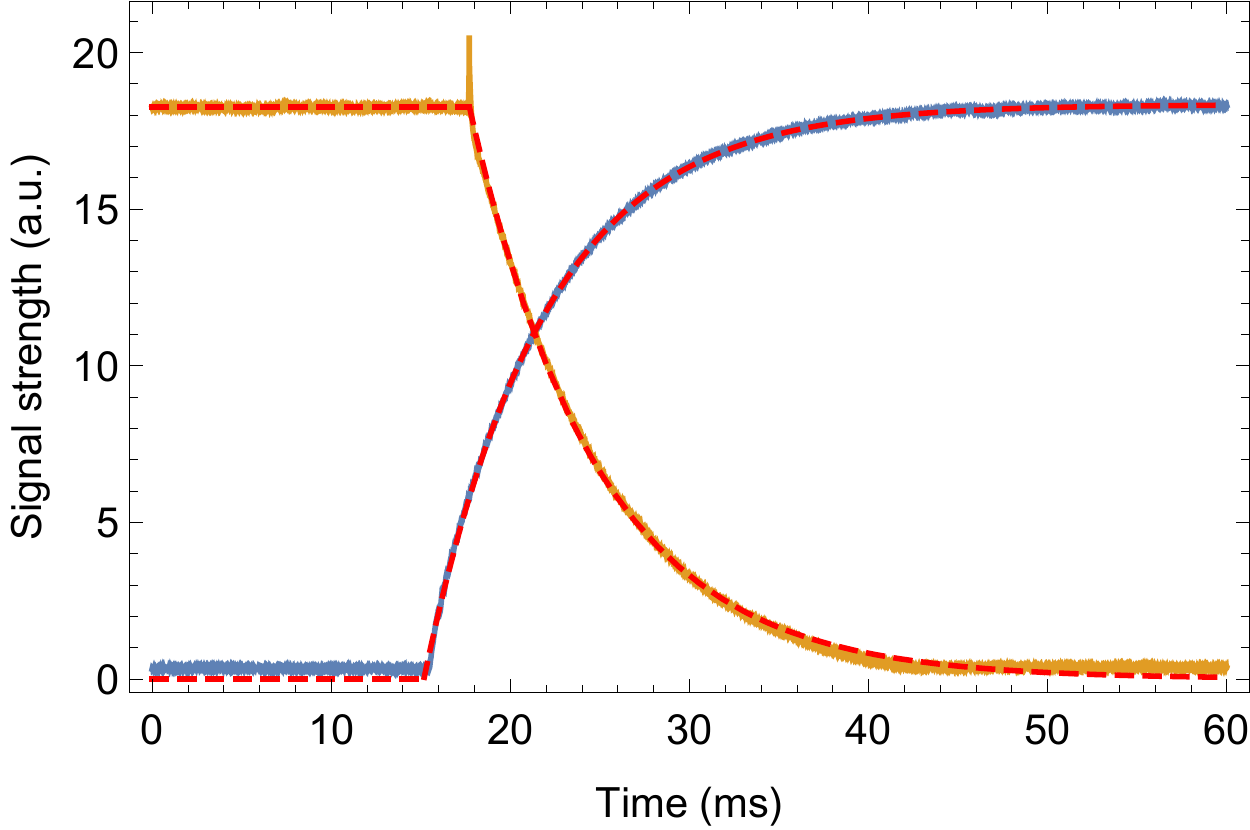}
	\caption{Oscilloscope traces corresponding to the exponential rising and falling edges for the current pulsed through the prepolarizer when driven with the IPM-16P and battery bank. The coil current in this example is $I_\text{p}\approx\SI{255}{A}$. The red dashed lines are fits to simple exponential functions, yielding $\tau\approx\SI{6.7}{ms}$ (\SI{7.2}{ms}) for the rising (falling) edge, consistent with expectations (see text). The spike right before the rising edge is a fast transient due to the switching electronics in the IPM-16P.}
	\label{fig:Pulses}
\end{figure}

Figure~\ref{fig:BatterySetup} also shows an inductive probe for current measurement (Danisens DS600ID), which we use to monitor the system's performance and characterize the prepolarizing pulse properties. The only high power active element in the system is the IPM-16P module, so the on-off transitions at the coil are voltage controlled and the current intensity shows an exponential (rather than linear) behavior, with an expected time constant $\approx\SI{6.4}{ms}$ when a battery pair delivers 24~V. The structure of the measured pulses can be see in Fig.~\ref{fig:Pulses} and is consistent with the above expectations.

\section{Installation into ``DentMRI - Gen I''}
\subsection{Mechanical installation}

\begin{figure}
	\centering
	\includegraphics[width=1.\columnwidth]{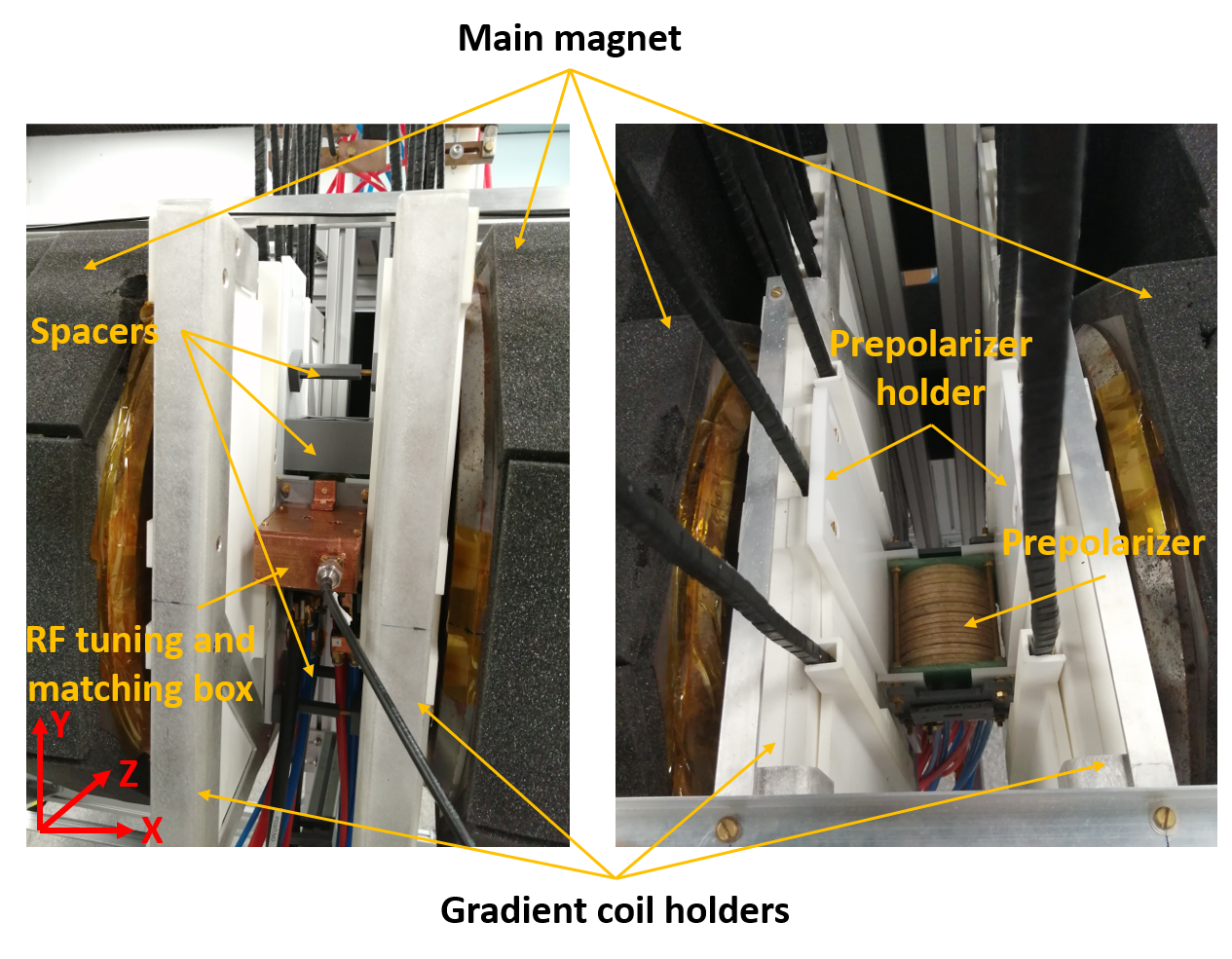}
	\caption{Front (left photograph) and top (right photograph) views of the prepolarizer module after installation in the ``DentMRI - Gen I'' scanner. The front view shows the spacers used for mechanical stability (see text), as well as a box containing tuning and matching electronics between the high power rf amplifier and the resonant coil. The prepolarizer magnet and holder are best viewed on the right photograph, which was taken right after the module was installed.}
	\label{fig:PrepolInt}
\end{figure}

The prepolarizer module is bolted to the nylon holder that accommodates the planar gradient coils of the ``DentMRI - Gen I'' system \cite{Algarin2020}. As shown in the photograph in Fig.~\ref{fig:PrepolInt}, the prepolarizer coil axis is orthogonal to the main field and in the plane normal to the magnet poles, which facilitates the insertion and extraction of the rf coil and sample through the main opening of the C-shaped permanent magnet. All electrical and hydraulic connections are on the bottom.

An important mechanical consideration is the intense Lorentz force that appears between the prepolarization and evolution magnets when the prepolarizer is on. Due to their orientation, the interaction between the magnets results in a torque on the prepolarizer module, generating a moment that points downwards. Since the prepolarization field is applied in long pulses (typically >500~ms to saturate the sample magnetization to the total field $B_\text{tot}=(B_0^2+B_\text{p}^2)^{1/2} \approx \SI{0.55}{T}$), the mechanical stress is also pulsed. This is problematic if the prepolarizer is not tightly fixed, since the whole module (including the rf coil and the sample) shifts to a different place during the application of the prepolarization field, and the sudden return to its relaxed position results in sample motion which compromises the quality of MR images. To address this, we have 3D-printed and installed a number of adjustable spacers which press against the inner surfaces of the prepolarizer and gradient coil holders (see Fig.~\ref{fig:PrepolInt}). In this way, we suppress mechanical motion to the point where we cannot observe any influence on the detected rf signals.

\subsection{Noise considerations}\label{sec:noise}

\begin{figure}
	\centering
	\includegraphics[width=1.\columnwidth]{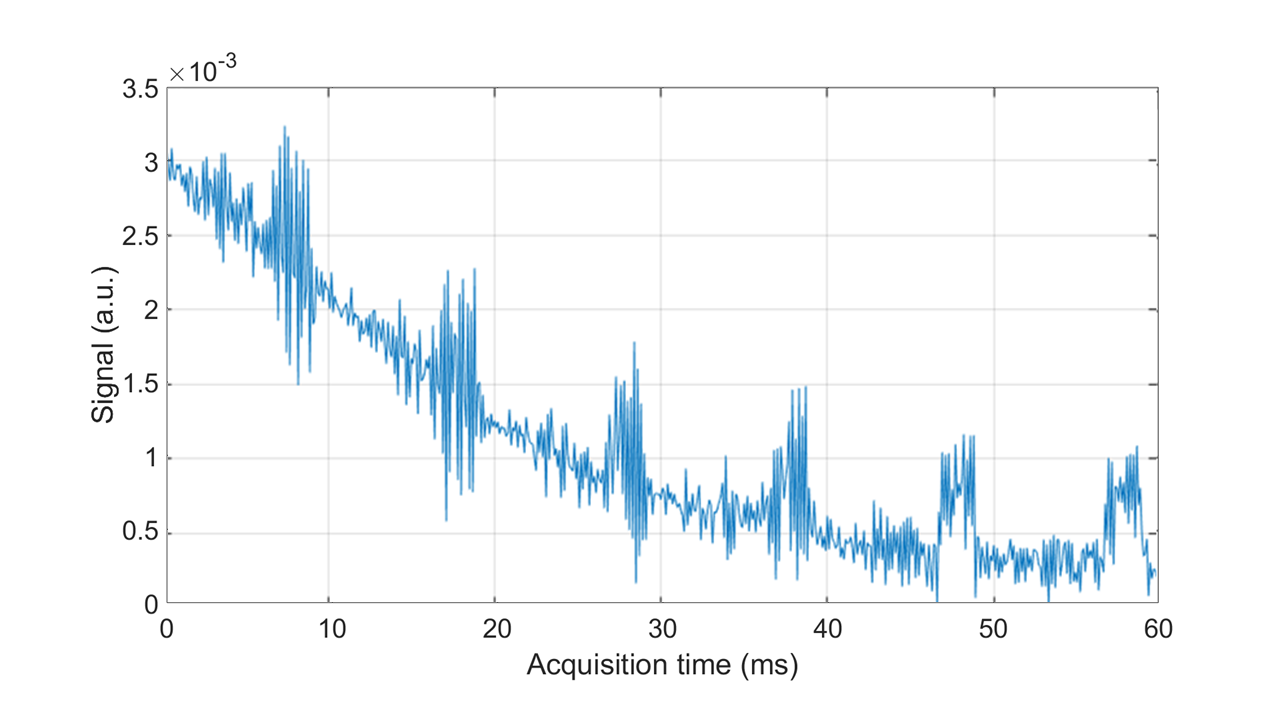}
	\caption{Free Induction Decay of a mouse brain sample before rf grounding of the prepolarizer cables' shielding. The overall noise level decreases significantly after proper grounding, and the periodic bursts every 10~ms (presumably the second harmonic from the 50~Hz network) disappear.}
	\label{fig:FIDNoise}
\end{figure}

Another relevant aspect is to prevent noise present in the laboratory (or generated by our electronics) from being picked up by the prepolarizer electrical connections and affecting the weak rf signal which we use for image reconstruction in MRI. To this end, scanners invariably use Faraday isolating cages around the resonant detection coils. With the prepolarizer module installed, the Faraday cage is inside the prepolarizer, directly around the small rf coil, which is coaxial to the prepolarization module. We found this necessary but insufficient when the thick cables from the IPM-16P outputs are connected to the prepolarizer load (Sec.~\ref{sec:elec}). This was solved by shielding the cables and connecting them to the rf ground. For reference, Fig.~\ref{fig:FIDNoise} shows a Free Induction Decay (FID, \cite{BkHaacke}) curve when unshielded.

\section{Prepolarized free induction decay curves}

\begin{figure*}
	\centering
	\includegraphics[width=2.\columnwidth]{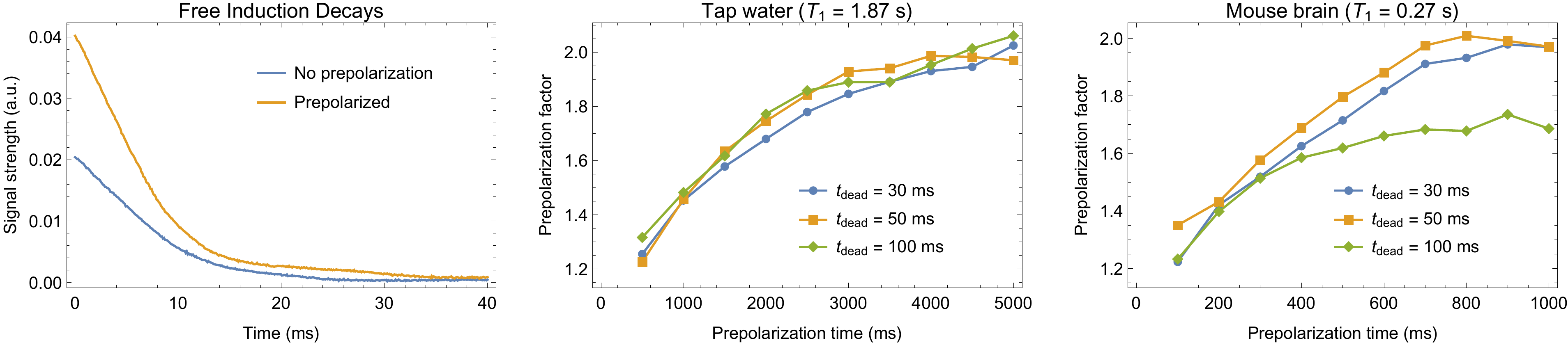}
	\caption{(a) FIDs of a sample containing tap water without and with prepolarization ($t_\text{p}=4$~s, $B_\text{tot} \approx \SI{0.55}{T}$, $t_\text{dead}=\SI{50}{ms}$). At $t=0$ we measure $\alpha\approx2$, in agreement with theoretical expectations (see text). (b) Prepolarization factor $\alpha$ measured for tap water as a function of the prepolarization time $t_\text{p}$, for $t_\text{dead}=30$, 50 and \SI{100}{ms}. (c) Idem for a mouse brain.}
	\label{fig:FIDs}
\end{figure*}

In order to test the performance of the prepolarizer module, we introduce tap water as a sample. For a first check, we run a simple sequence of two pulses: a long prepolarization pulse ($t_\text{p}=4$~s and $B_\text{tot} \approx \SI{0.55}{T}$), and a $\pi/2$ resonant rf pulse to coherently rotate the sample magnetization to the transverse direction. Both pulses are separated by a wait time $t_\text{dead}=\SI{50}{ms}$, long enough to largely remove the prepolarizing field ($\tau\approx\SI{6.4}{ms}$), which otherwise shifts the Larmor frequency and also distorts the FID signal due to its intrinsic inhomogeneity. The FID is acquired for 40~ms after the rf pulse. Figure~\ref{fig:FIDs}(a) shows the absolute value of the FID (after quadrature demodulation at $\omega_0$) with and without prepolarization pulse. We define the prepolarization factor $\alpha$ as the ratio between the initial amplitudes of these curves, and it is expected to be:
\begin{equation}\label{eq:alpha}
	\alpha(t_\text{dead},T_1) = \alpha_0\Exp{-t_\text{dead}/T_1},
\end{equation}
with
\begin{equation}\label{eq:alpha0}
	\alpha_0(t_\text{p},T_1) = 1+\frac{B_\text{tot}-B_0}{B_0}(1-\Exp{-t_\text{p}/T_1})
\end{equation}
the prepolarization factor right after the prepolarizing pulse, where $B_\text{tot}=(B_0^2+B_\text{p}^2)^{1/2}$ in our system due to the geometric arrangement. We have measured the $T_1$ of the tap water to be $\approx\SI{1.87}{s}$ using a simple Inversion Recovery sequence \cite{BkHaacke}. This means we expect $\alpha\approx2.05$, in good agreement with the increase in polarization observed in the plots.

To complete the characterization of the system, we measured $\alpha$ as a function of the prepolarization pulse length $t_\text{p}$. This is shown in Fig.~\ref{fig:FIDs}(b) for water, and in Fig.~\ref{fig:FIDs}(c) for a fixed mouse brain (further details in Sec.~\ref{sec:images}), both of which are consistent with the above model. The data points in this plot are calculated as the ratio of the FID starting amplitudes with and without prepolarization. As expected, the dead times used have little influence on the water sample due to its comparatively long $T_1$, whereas $\alpha$ is smaller for longer wait times with the mouse brain.

\section{Prepolarized magnetic resonance images}\label{sec:images}

\begin{figure*}
	\centering
	\includegraphics[width=2.\columnwidth]{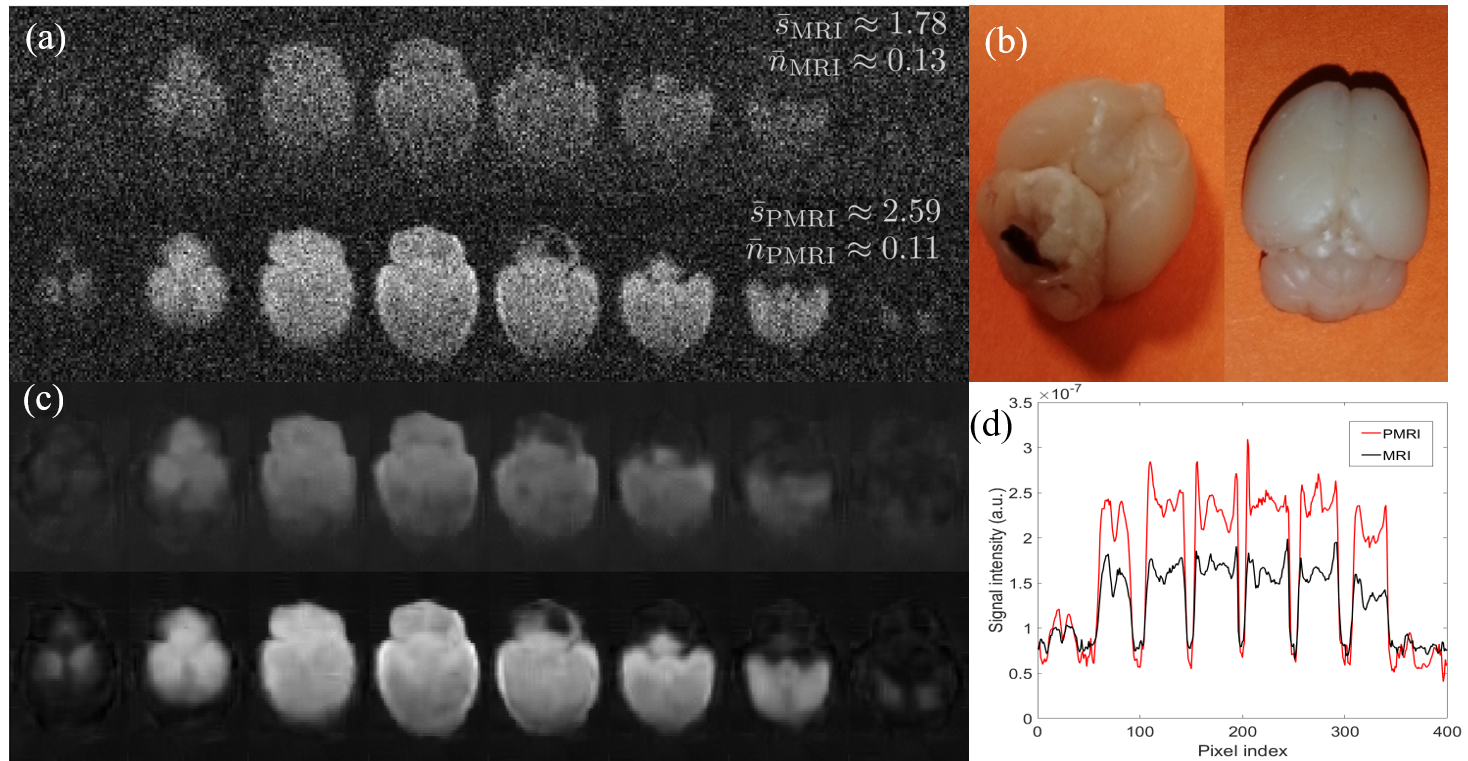}
	\caption{(a) MRI (top) and PMRI (bottom) of an \emph{ex-vivo} mouse brain, imaged with a RARE pulse sequence (see Tab.~\ref{tab:Pulses}). (b) Photographs of the brain after extraction and tissue fixation. (c) Same as (a) after denoising with a Block-Matched Filter. (d) Signal intensity along the middle horizontal lines of the images in (a). The curves in (d) are low-pass filtered to aid visual estimation of the SNR increase due to prepolarization. The value obtained for the prepolarization gain ($\alpha\approx1.75$) is in excellent agreement with the expected value of 1.748 (see text).}
	\label{fig:Brain}
\end{figure*}

The final goal of the prepolarizer module is to enhance the SNR in the reconstructed images. To this end, we obtain \emph{ex vivo} images of brain samples extracted from OF-1 albino, 2 to 4-month-old mice (see Fig.~\ref{fig:Brain}(b)). The animals were anesthetized with an intraperitoneal dose of sodium pentobarbital (60 mg/kg) and later perfused with a fixative solution composed of 4~\% paraformaldehyde in 0.1~M phosphate buffer, pH 7.2. The brain, after being removed, was immersed overnight in the same fixative to prolong the useful lifetime of the sample. Prior to imaging, the $T_1$ of the heterogeneous, fixed brain is measured in the scanner by Inversion Recovery, yielding $\approx\SI{270}{ms}$.

\begin{table}
	\caption{Prepolarization and pulse sequence parameter values for the images in Figs.~\ref{fig:Brain} and \ref{fig:Contrast}, using high power electronics based on commercial boat batteries. The rise, fall and wait times expected with the Danfysik pulsed power supply unit are included at the bottom.}
	\centering
	\begin{tabular}{c c c}
		\hline
		Parameter & Brain PMRI & Contrast PMRI\\
		 & (Fig.~\ref{fig:Brain}) & (Fig.~\ref{fig:Contrast})\\
		\hline
		Prepol. field ($B_\text{p}$) & \SI{0.48}{T} & \SI{0.41}{T}\\
		Pulse intensity ($I_\text{p}$) & \SI{255}{A} & \SI{214}{A}\\
		Flat top ($t_\text{p}$) & \SI{700}{ms} & \SI{700}{ms}\\
		Voltage ($V$) & \SI{24}{V} & \SI{20.4}{V}\\
		Rise\&fall time ($\tau$) & \SI{6.4}{ms} & \SI{6.3}{ms}\\
		Wait time ($t_\text{dead}$) & \SI{40}{ms} & \SI{40}{ms}\\
		Spin relax. time ($T_1$) & \SI{0.27}{s} & Not measured\\
		Prepol. gain ($\alpha$) & 1.75 & Various\\
		\hline
		Echo time (TE) & \SI{5}{ms} & \SI{5}{ms}\\
		Echo train len. (ETL) & 2 & 2\\
		Repetition time (TR) & \SI{800}{ms} & \SI{1000}{ms}\\
		Field of View (FoV) & $14\times10\times\SI{9}{mm^3}$ & $28\times12\times\SI{14}{mm^3}$\\
		Number of voxels & $70\times50\times8$ & $56\times1\times28$\\
		Voxel size & $0.2\times0.2\times\SI{1.1}{mm^3}$ & $0.5\times12\times\SI{0.5}{mm^3}$\\
		Acq. bandwith (BW) & $2\pi\cdot\SI{23.3}{kHz}$ & $2\pi\cdot\SI{23.3}{kHz}$\\
		Number of averages & 33 & 5\\
		Total acq. time ($T_\text{acq}$) & $\SI{90}{min}$ & $\SI{1}{min}$\\
		\hline
		Max. voltage w. PPSU ($V$) & \SI{500}{V} & \SI{500}{V}\\
		Rise\&fall time w. PPSU ($\tau$) & \SI{300}{\micro s} & \SI{300}{\micro s}\\
		Wait time w. PPSU ($t_\text{dead}$) & <\SI{1}{ms} & <\SI{1}{ms}\\
		\hline
	\end{tabular}
	\label{tab:Pulses}
\end{table}

The images in Fig.~\ref{fig:Brain}(a) and (c) are obtained with a Rapid Acquisition with Relaxation Enhancement (RARE) pulse sequence \cite{Hennig1986}, using an echo train length of 2. The size of the field of view is set to $14\times10\times\SI{9}{mm^3}$ and the image is reconstructed by a Fast Fourier Transform (FFT) protocol into $70\times50\times8$ voxels. The echo spacing and repetition time are set to $\text{TE}=\SI{5}{ms}$ and $\text{TR}=\SI{800}{ms}$, with a bandwidth of $\approx\SI{23}{kHz}$. $k$-space data were acquired on a low-high linear trajectory, as in Ref.~\cite{OReilly2020}. Every image contains 33 averages for a total scan time of $\approx\SI{90}{min}$ (further details in Tab.~\ref{tab:Pulses}). The bottom rows of images in Fig.~\ref{fig:Brain}(a) and (c) correspond to scans in which a prepolarization pulse of 700~ms at $B_\text{p}\approx\SI{0.48}{T}$ is applied at the beginning of a sequence repetition, and where $t_\text{dead}=\SI{40}{ms}$. The pulse sequence for the top rows of images is identical, but the prepolarization pulse is not triggered ($B_\text{p}=0$). The color scale is common to both datasets to highlight the gain in SNR when we prepolarize the sample. The fixation process changes the physical properties of the sample and hampers contrast between the different brain tissues, especially at low magnetic field strengths \cite{Kanawaku2014}.

Figure~\ref{fig:Brain}(a) shows the raw images with MRI and PMRI, which we can Block-Match filter \cite{Algarin2020,Maggioni2013} for a cleaner result (Fig.~\ref{fig:Brain}(c)). The SNR enhancement is evident in both sets of images. In order to quantify the influence of prepolarization, we plot in Fig.~\ref{fig:Brain}(d) the signal intensity profile along a horizontal line around the middle portion of the images in (a). The results shown are low-pass filtered (in image space) to aid visual estimation of the SNR increase. The noise levels are almost identical with and without prepolarization, as can be seen in the images, so this filtering does not mask noise effects. The mean $\alpha=\text{SNR}_\text{PMRI}/\text{SNR}_\text{MRI}$ (averaged over a region of interest of bright pixels around the brain center) is $\approx1.72$, where $\text{SNR}_\text{PMRI}=\bar{s}_\text{PMRI}/\bar{n}_\text{PMRI}\approx23.6$, and $\text{SNR}_\text{MRI}$ (analogously defined) is $\approx13.7$. The mean signal and noise values ($\bar{s}$ and $\bar{n}$, see Fig.~\ref{fig:Brain}(a)) are estimated, respectively, as the mean and standard deviation of the voxel brightness in the region of interest. For comparison, the expected prepolarization gain from Eqs.~(\ref{eq:alpha}) and (\ref{eq:alpha0}) is $\approx1.75$ with the values in Tab.~\ref{tab:Pulses} ($T_1\approx\SI{0.27}{s}$).

\begin{figure*}
	\centering
	\includegraphics[width=1.75\columnwidth]{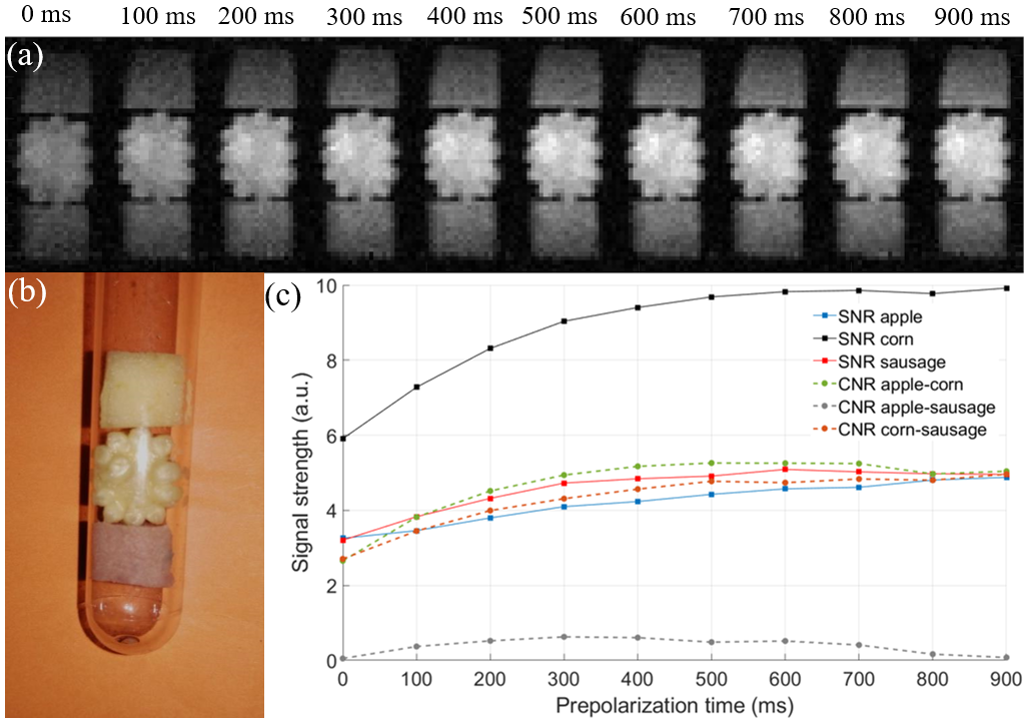}
	\caption{(a) MRI (0 ms) and PMRI (rest) of an organic sample consisting of bits of apple (top), baby corn (middle) and sausage (bottom), imaged with a RARE pulse sequence (see Tab.~\ref{tab:Pulses}). (b) Photograph of the sample. (c) SNR of the different sample components and CNR between pairs of them, as a function of the prepolarization time.}
	\label{fig:Contrast}
\end{figure*}

Since the distinguishability between fixed brain tissues is low, we use a different sample to demonstrate tissue contrast in our PMRI setup. The sample consists of parts of apple, baby corn and sausage (Fig.~\ref{fig:Contrast}(b)). The images reconstructed in Fig.~\ref{fig:Contrast}(a) are taken for a RARE PMRI sequence as in Fig.~\ref{fig:Brain}, where the size of the field of view is set to $28\times12\times\SI{14}{mm^3}$ and the (projection) image is reconstructed into $56\times1\times28$ voxels. The echo spacing and repetition time are set to $\text{TE}=\SI{5}{ms}$ and $\text{TR}=\SI{1}{s}$, with a bandwidth of $\approx\SI{23}{kHz}$. Every image contains 5 averages for a total scan time of $\approx\SI{1}{min}$ (further details in Tab.~\ref{tab:Pulses}). The prepolarization times are shown above the corresponding images, ranging from no prepolarization to 900~ms, in steps of 100~ms. Variations in the proton densities and relaxation and dephasing times of the various compounds result in different SNRs (see Fig.~\ref{fig:Contrast}(c)). In particular, the growth rate of the SNR as a function of $t_\text{p}$ is mostly dependent on $T_1$. Consequently, the contrast-to-noise ratio (CNR) between pairs of tissues, defined as the absolute value of the difference between their SNRs, can be tuned with $t_\text{p}$, enabling $T_1$ contrast with PMRI \cite{Matter2006-2}. The evolution of CNRs as a function of the prepolarization pulse length is also shown in Fig.~\ref{fig:Contrast}(c). We find that, with the TE and TR settings employed there is no contrast between the apple and the sausage, whereas prepolarizing the sample leads to significant contrast, especially between baby corn and sausage/apple.

\section{Conclusion and outlook}
In conclusion, we have designed, built and characterized a prepolarization module that can be easily incorporated into our ``DentMRI - Gen I'' scanner. The reduced size of the field of view allows for intense prepolarization (up to 0.7~T with the current cooling system) while opening the door to sub-milli-second switching. This makes the prepolarizer suitable for use with solid-state samples and hard biological tissues. These have short $T_1$ times, especially for weak evolution fields, and they remain out of reach for standard PMRI.

With this work, we have demonstrated the advantages of prepolarization on long $T_1$ samples, with high power electronics previously available to us. Although they are not specifically designed for this system, they allow the prepolarizer to be switched on and off in tens of milli-seconds for $B_\text{p} \approx \SI{0.5}{T}$. This is comparable to the fastest prepolarizers reported \cite{Matter2006}, but not fast enough for efficient SNR boost in short $T_1$ samples. Nevertheless, the observed increase in signal-to-noise ratio is in excellent agreement with our models for the electronics and time evolution of the magnetization. This suggests that more advanced electronics, which we are already working on, can shunt the prepolarization field in $\sim\SI{300}{\micro s}$, much faster than required for hard tissue PMRI.

The setup presented in this work is suitable for experimental and preclinical operation (with e.g. rodents), given its size. However, the electronics under development will deliver up to 500~V. This is sufficient to drive a $\sim\SI{10}{mH}$ coil, which can be large enough for human extremity imaging \cite{Matter2006}, in $\sim\SI{5}{ms}$.

\section*{Contributions}

The prepolarizer was designed, assembled and characterized by JPR, CG and EP, with contributions from DGR and JA. Experimental data in the ``DentMRI - Gen I'' scanner were taken by JB, JMG and JPR, with help from JMA, RB, FG, RP and JA. Data analysis performed by JA, JPR, JB and JMG, with input from JMA, FG and RP. Animal handling and manipulation of biological tissues performed by JVSA. The paper was written by JA, JPR and JB with input from all authors. Experiments conceived by JMB, JA and AR.


%

%

\section*{Acknowledgment}
This work was supported by the European Commission under grant 737180 (FET-Open: HistoMRI) and Ministerio de Ciencia e Innovaci\'on of Spain for research grant PID2019-111436RB-C21. Action co-financed by the European Union through the Programa Operativo del Fondo Europeo de Desarrollo Regional (FEDER) of the Comunitat Valenciana 2014-2020 (IDIFEDER/2018/022). JMG acknowlegse support from the Innodocto program of the Agencia Valenciana de la Innovación (INNTA3/2020/22). JVSA acknowledges support from UJISABIO through grant number 21I044.01/1.

\section*{Ethical statement}
The experiments were carried out according to institutional animal care guidelines. Animal housing and all protocols were approved and in accordance with institutional guidelines, Spanish animal protection laws, and conformed to the Directive 2010/63 EU of the European Parliament. The experimental protocol was approved by the Ethics Committee for Research and Animal Welfare of the University Jaume I (License reference: 2019/VSC/PEA/0132). Mice were housed in standard conditions.


\end{document}